\documentclass[12pt]{article}
\usepackage{epsfig}

\usepackage{cite}

\def\Title#1{\begin{center} {\Large {\bf #1} } \end{center}}

\begin{document}

\Title{Naturalness after LHC8}

\bigskip\bigskip


\begin{raggedright}  

{\it Gian Francesco Giudice\index{Giudice, G.F.}\\
CERN, Theory Division\\
1211 Geneva 23, Switzerland}\\
\bigskip
Talk delivered at the 2013 Europhysics Conference on High Energy Physics (EPS), Stockholm, Sweden, 18-24 July 2013.
\bigskip\bigskip
\end{raggedright}

The main task of the LHC was to clarify the mechanism of electroweak symmetry breaking. Now that the Higgs boson has been discovered, the most pressing goal is to elucidate the origin of this particle and, in this respect, naturalness has become the central issue. Since long ago~\cite{wilson,gildener,weinberg,thooft}, it had been recognised that the Higgs boson introduces a conceptual problem associated with the quantum corrections to its mass term. In an effective-theory approach where momenta of virtual particles are cut off at the scale $\Lambda$, the quantum corrections to the physical Higgs mass grow proportionally with $\Lambda$
\begin{equation}
\label{corr}
\frac{\delta m_h^2}{m_h^2}=\frac{3G_F}{4\sqrt{2}\pi^2}\left( \frac{4m_t^2}{m_h^2}-\frac{2m_W^2}{m_h^2}-\frac{m_Z^2}{m_h^2}-1\right)
\Lambda^2=\left(\frac{\Lambda}{500~{\rm GeV}}\right)^2\ .
\end{equation}
This is generally taken as an indication that a simple extrapolation of the Standard Model (SM) beyond a scale of about 500~GeV suffers from a naturalness problem because ultraviolet (UV) contributions to $m_h^2$ exceed its physical value. The larger the value of $\Lambda$, the more acute the problem becomes, although the maximum acceptable $\Lambda$ is a matter of subjective judgement. (For general discussions about naturalness, see refs.~\cite{Giudice,Feng,Wells}.)  

The expectations for new physics beyond the SM have not been substantiated by the first 20~fb$^{-1}$ of data from the LHC operating at 8~TeV. There is no doubt that these experimental results represent a considerable challenge to the idea that naturalness of the electroweak breaking is restored by new dynamics below the TeV. This has led to an intense debate inside the high-energy physics community about the validity of the naturalness principle. A commonly asked question is: 

\section*{Is the effective field-theory approach unreliable?}

 In other words: is it misleading to give a physical interpretation to the UV cutoff in eq.~(\ref{corr})? My answers to this question is a definite no. But to avoid getting confused with regularisation procedures in effective theories and with the physical meaning of a cut-off parameter, it is always best to frame the naturalness problem in a setup where we replace $\Lambda$ by an explicit particle mass scale, widely separated from the weak scale. (Arguing whether the pure SM is natural or not is an ill-posed question because, in the presence of a single mass scale, the naturalness problem cannot even be formulated.) If we add to the SM potential $V=m_H^2|H|^2+\lambda |H|^4$ a new scalar field $\Phi$ with a large mass $M$ ($M\gg m_H$) and an interaction $\lambda_\Phi |H|^2 |\Phi|^2$, the electroweak (EW) scale is destabilised by a logarithmically-divergent contribution 
\begin{equation}
\label{gut}
\delta m_H^2 \approx \frac{\lambda_\Phi}{16\pi^2} M^2 \ln \frac{M^2}{\Lambda^2}+\dots 
\end{equation}
This is exactly what happens in traditional GUT models, where $\Phi$ is a field associated with the breaking of the unified gauge group~\cite{gildener,weinberg}. This  example shows that the occurrence of the naturalness problem is unrelated to regularisation issues associated with power divergences. This is even more evident in the case of a supersymmetric extension of the SM, in which supersymmetry is broken by a very large stop mass ${\tilde m}_t$ (${\tilde m}_t \gg m_H$). The theory is free from quadratic divergences, and yet the EW scale is badly destabilised by terms of the form
\begin{equation}
\label{susy}
\delta m_H^2 = \frac{3y_t^2}{8\pi^2} {\tilde m}_t^2 \ln \frac{{\tilde m}_t^2}{\Lambda^2}+ \dots \ ,
\end{equation}
where $y_t$ is the top Yukawa.

These simple examples illustrate some important (and well-known) features about the issue of Higgs naturalness. The naturalness problem appears whenever new massive states, whose mass terms are invariant under the EW gauge group, are coupled to the Higgs field. This is because $m_H$ is renormalised {\it additively} (as opposed to {\it multiplicatively}), so that quantum corrections are parametrically uncorrelated with the classical value of $m_H$ and can be numerically much larger. In turn, the additive renormalisation comes from the fact that there is no symmetry enhancement in the limit $m_H \to 0$, as emphasised long ago by 't Hooft~\cite{thooft}. Naively, one could think that, in the limit in which the Higgs quadratic term is set to zero, the theory acquires a new conformal symmetry, at least at the classical level. However, as mentioned before, the naturalness problem can be meaningfully formulated only in the presence of a mass scale $M$ much larger than $m_H$. If there is a large mass separation, conformal symmetry is badly broken by $M$. If there is no large mass separation, we have no naturalness problem to start with. As a result, in this context, conformal symmetry is of no avail to address the naturalness problem. (We will see later a case in which conformal symmetry could have something to do with naturalness.) The other lesson we have learned from the previous examples is that the naturalness problem of the Higgs is completely insensitive to the regularisation procedure. The vanishing of quadratic divergences in dimensional regularisation has no bearing on the problem.

The next question is: 

\section*{Is naturalness a good guiding principle?}

The naturalness principle is certainly not a {\it necessary} condition, indispensable for the internal consistency of the theory. However, it is also not a purely {\ae}sthetic requirement. It is the consequence of a reasonable criterion that assumes the lack of special conspiracies between phenomena occurring at very different length scales. It is deeply rooted in our description of the physical world in terms of effective theories. Separation of scales is not an {\it a priori} necessary ingredient, but it has been a cornerstone of much of the progress done in physics throughout the centuries. Were it necessary for deriving the trajectory of the Moon's orbit to solve the equation of motion of each individual quark and electron in the lunar interior, how could have Newton obtained his gravity equation? Separation of scales has been a very useful tool for physicists to make progress along the path towards the inner layers of matter, and we can be grateful to nature for employing it so generously. Indeed, one can find numerous examples in which the naturalness principle applied to an effective theory gives (or -- to be historically correct -- could have given) the right hint for the existence of energy thresholds at which there are changes in the physical degrees of freedom describing a certain phenomenon. I have already discussed elsewhere (ref.~\cite{Giudice}, p.~12--15) how the electromagnetic energy of a classical electron poses a naturalness problem that is cured by the positron;  how the naturalness of the pion mass difference is cured by the coming of the rho meson; and the naturalness of the neutral kaon mass difference by the charm quark. I will not repeat these arguments here. The lesson to be learnt from these examples is that naturalness successfully works as a warning signal for the presence of a new layer in the stack of effective theories describing nature.  

As already stated above, naturalness is not a necessary condition. This was painfully learnt by physicists through the observation of dark energy. The related cosmological constant corresponds to an energy scale of $2.4 \times 10^{-3}$~eV and the naturalness principle predicts a new-physics threshold below this scale. The lack of empirical evidence for such a threshold gives support to the conjecture that dark energy does not respect the naturalness principle.

The LHC is now scrutinising the situation regarding the naturalness principle in the case of the Higgs boson. At this stage, theory is unable to provide a definite answer and the issue has become largely an experimental question, in which every outcome is possible. In this investigation there is more at stake than a quibble among theoreticians about an abstract criterion or than the properties of a single particle -- the Higgs boson.
The outcome of Higgs naturalness will be decisive for the future of particle physics. Not only will it determine future research directions in theoretical physics, but it will forcefully influence the experimental strategies to be pursued by high-energy experimental physics. Given the uncertainty and the importance of the issue, it is a good moment to consider all various possible outcomes with an open mind. 
 
\section*{Unnaturalness} 

It is conceivable that the LHC will find that the Higgs mass does not respect the naturalness criterion, just like (probably) the case of the cosmological constant. Accepting this possibility, however, does not imply that we can simply ignore the issue. As I argued above, naturalness is a well-posed problem, deeply rooted in our understanding and formulation of effective theories. In other words: if we accept {\it Unnaturalness}, we have to address the question of why the Higgs is unnatural. 

At the moment, the multiverse offers the most plausible answer at our disposal. The landscape of string vacua together with the dynamical process of eternal inflation provides a reasonable theoretical setup for the multiverse.  
In the multitude of universes, anthropic arguments find the necessary statistical ensemble to explain the `unnatural' sizes of both the cosmological constant~\cite{wei} and the weak scale~\cite{don}. In spite of these virtues, the multiverse remains the option most hated by the vast majority of physicists. It is curious how the idea is often rejected using emotional arguments, which are as unscientific as the alleged unscientific character of the multiverse.
But, leaving aside philosophical convictions, the common fear is that the idea of the multiverse is doomed to be experimentally untestable. There is no doubt that probing universes that are causally disconnected appears to be a very arduous experimental task and we can only hope for some future clever idea. (By the way, when inflation was first proposed, it also looked like a good, but untestable, proposal. The idea of measuring quantum fluctuations stretched to super-horizon scales came later...) 

\begin{figure}[htb]
\begin{center}
$$\includegraphics[width=0.45\textwidth]{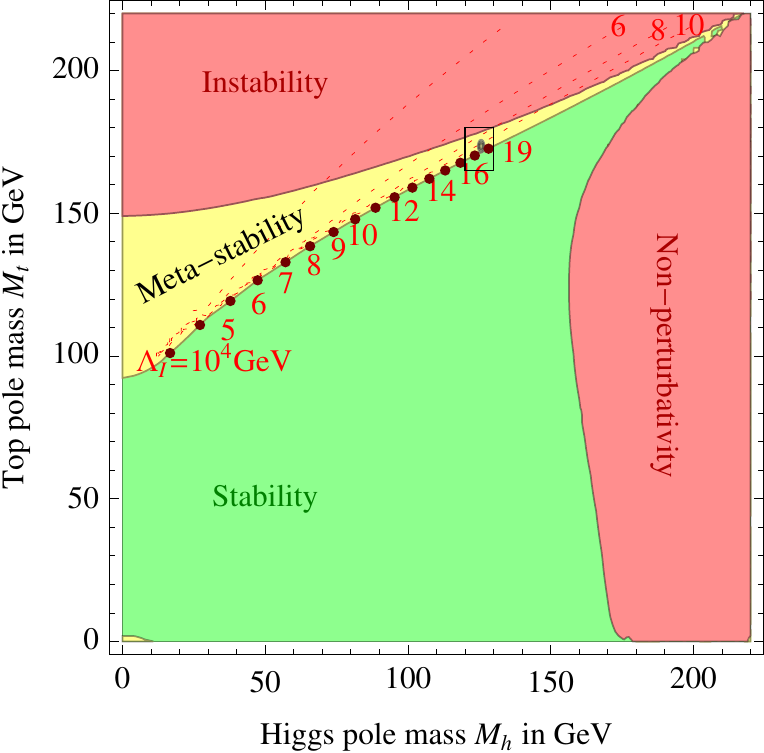}\qquad
\includegraphics[width=0.45\textwidth]{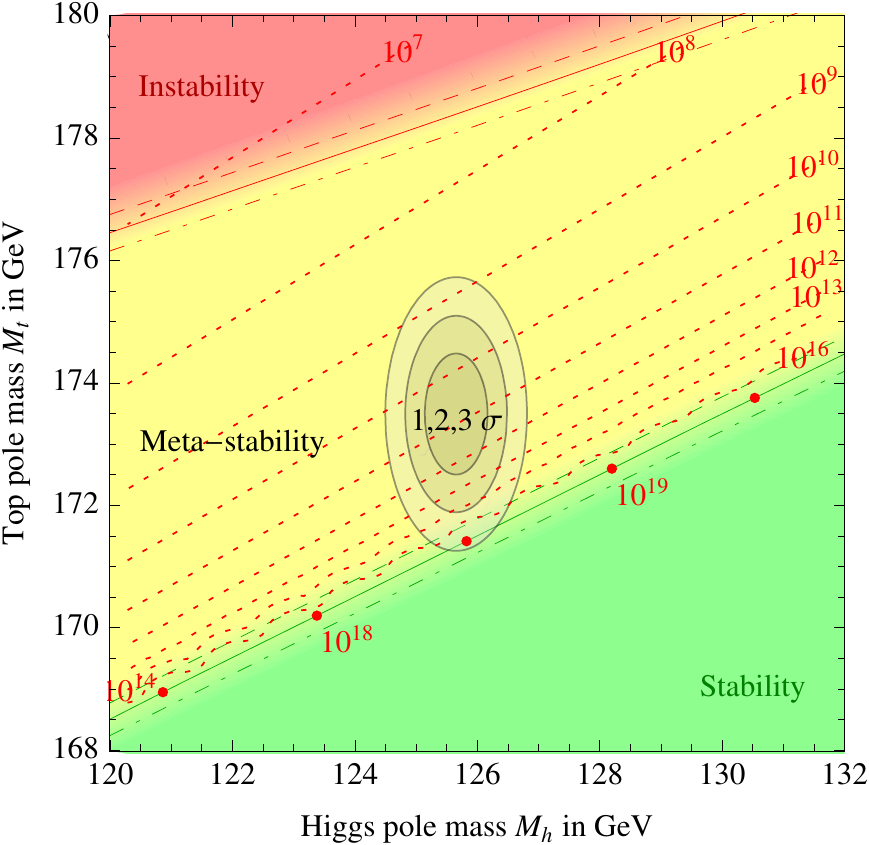}$$
\caption{Regions of absolute stability, meta-stability and instability of the SM vacuum 
in terms of the top and Higgs masses.
The frame on the right zooms into the preferred experimental region 
(the grey ellipses denote the allowed region at 1, 2, and 3$\sigma$).
The three  boundary lines correspond to $\alpha_s(M_Z)=0.1184\pm 0.0007$, and the grading of the 
colours indicates the size of the theoretical error.  
The dotted contour-lines show the instability scale in GeV, assuming the central value of $\alpha_s(M_Z)$. (For details see refs.~\cite{Degrassi:2012ry,Buttazzo:2013uya}.)}
\label{fig:regions}
\end{center}
\end{figure}

For the moment, we must content ourselves to look for peculiarities in some of the measured parameters. In this respect, the Higgs potential is a particularly rich source of information:
\begin{equation}
\label{higpot}
 V={\rm const.} + m_H^2|H|^2+\lambda |H|^4 \ .
\end{equation}
The measured values of all three parameters in eq.~(\ref{higpot}) raise theoretical concerns and turn out to be close to `living dangerously' conditions. The constant in eq.~(\ref{higpot}) (which actually represents the full vacuum energy of the theory) leads to the cosmological constant problem and $m_H^2$ embodies the problem of Higgs naturalness. The recent measurement of the Higgs boson mass has added a new ingredient to the list: $\lambda$ is special with respect to vacuum stability~\cite{Degrassi:2012ry,Buttazzo:2013uya,Bezrukov:2012sa}. Small variations of any of these three parameters with respect to their measured values could have devastating consequences for our life-friendly universe. The situation for $\lambda$ is quantified in fig.~1 (for details about this figure, see refs.~\cite{Degrassi:2012ry,Buttazzo:2013uya}). If the SM is assumed to be valid up to very short distances (much beyond what can be tested experimentally today), the measured Higgs boson mass corresponds to a near-critical situation in which the EW vacuum is on the verge of instability.

Unnaturalness does not mean that there is nothing left to discover. Beside the Higgs naturalness problem, there are still many open questions in particle physics and each of them could imply the existence of new physics. The origin of flavour-symmetry breaking, dark matter, the strong CP problem, baryogenesis, inflation, unification of gauge forces, dark energy, charge quantisation can all be viewed as valid motivations for new particles and new phenomena. However, while naturalness strictly guarantees new discoveries at the LHC, it is not possible to link the other open problems of the SM to an energy scale which is {\it necessarily} within the LHC domain.

 It is curious that the supersymmetric model that (in my opinion) remains the most attractive possibility among the models that survived the brutal attack by LHC8 is, technically speaking, unnatural. I am referring to the model~\cite{Giudice:1998xp,welltemp} (see also refs.~\cite{Hall:2011jd,Bhattacherjee:2012ed,Arvanitaki:2012ps,Hall:2012zp,ArkaniHamed:2012gw}) in which squarks and sleptons get masses of the order of the gravitino mass, while gaugino masses arise through anomaly mediation~\cite{Randall:1998uk,Giudice:1998xp} and are a loop-factor smaller. This structure has several attractive theoretical and phenomenological 
features. It retains the positive aspects of split supersymmetry~\cite{split1,split2,split3,Wells:2003tf,Wells:2004di}
(gauge coupling unification, dark matter, ease of the flavour problem)
without requiring the artificial (although
possible~\cite{split1,split3,Luty:2002ff}) suppression of one-loop
anomaly-mediated gravitational contributions. It retains the positive
aspects of anomaly mediation (elegance, predictivity, viability of
dynamical supersymmetry breaking) without introducing the problem of
tachyonic sleptons~\cite{Randall:1998uk}. Moreover, it gives a prediction for the Higgs mass which is comfortably in the right range~\cite{Giudice:2011cg}, unlike most natural versions of supersymmetric models. Finally, it offers a chance for discovery at the high-energy phase of the LHC through gluino pair production, although it is not guaranteed that gluinos are kinematically accessible.

\section*{UV Naturalness}

As I have already mentioned, whenever we encounter a threshold with particles of mass $M$, coupled to the Higgs field, we expect that quantum corrections give a contribution 
 \begin{equation}
\label{nat}
 \delta m_H^2\approx \frac{\alpha}{4\pi}M^2 \ .
\end{equation}
This introduces a naturalness problem.

So let us suppose that no heavy particles coupled to the Higgs exist at all. For the moment I disregard all indications in favour of new heavy thresholds based on dark matter, strong CP, baryogenesis, inflation, unification, etc. Nonetheless, there is one mass scale I cannot dispense with: the Planck mass $M_{\rm Pl}$ associated with quantum gravity. This leads me to consider the following question: Does gravity introduce a Higgs naturalness problem? In practice, one would like to compute loop diagrams with two external Higgs lines, involving virtual gravitons and SM particles. Do these diagrams give a contribution  $\delta m_H^2\propto M_{\rm Pl}^2$ or not? In classical general relativity, the Planck mass enters only through the combination $G_N=M_{\rm Pl}^{-1/2}$, as a coupling with inverse powers of $M_{\rm Pl}$. Does quantum gravity introduce positive powers of $M_{\rm Pl}$ in the result? One generally expects that the answer is in the affirmative. Pure gravity loop diagrams do not contribute to the Higgs mass, because of the Higgs shift symmetry. But there is no obvious reason why two-loop diagrams involving gravity and top Yukawa (or Higgs quartic) couplings should vanish.
For instance, we can interpret microscopic black holes as virtual quantum states that contribute at the loop level to gravitational corrections $\delta m_H^2\propto M_{\rm Pl}^2$. However, since we cannot solve quantum gravity, it is difficult to make a firm statement. Some authors have considered (either implicitly or explicitly)~\cite{Bardeen:1995kv,Frampton:1999yb,Meissner:2006zh,Shaposhnikov:2007nj,Shaposhnikov:2009pv,Dubovsky:2013ira,Farina:2013mla,Iso:2013aqa,Chun:2013soa,Heikinheimo:2013fta,Hambye:2013dgv} the hypothesis that quantum gravity may not necessarily introduce any `Planckian particles' and quantum-gravity corrections to the Higgs mass may be free from positive powers of $M_{\rm Pl}$. Some (still unspecified) miracle is expected to cure the UV behaviour of gravity and the presence of $G_N$ would not significantly affect the Higgs mass.

Although it goes against effective field-theory intuition, one can conceive the peculiar possibility that quantum-gravity corrections $\delta m_H^2\propto M_{\rm Pl}^2$ vanish. It has never been proven to be true, but the opposite hasn't been proven either. This may not seem such a scientifically cogent reason, but it follows the same successful logic that Igor Stravinsky used, when he said ``Silence will save me from being wrong, but it will also deprive me of the possibility of being right." 

The basic observation is that quadratic divergences are fully related to UV physics. This means that, if the matching condition of the Higgs bilinear at an arbitrary scale $\Lambda$ in the far UV is $m_H(\Lambda)\approx 0$, then $m_H$ remains small at all scales below $\Lambda$, as long as there are no massive thresholds at intermediate energies. This is evident once we consider the one-loop renormalisation-group equation for $m_H$ in the SM
\begin{equation}
\label{rg}
\frac{d\, m_H^2}{d\, \ln \mu}=\frac{3m_H^2}{8\pi^2}\left( 2\lambda +y_t^2-\frac{3g_2^2}{4}-\frac{3g_1^2}{20}\right) \ .
\end{equation}
The Higgs parameter $m_H^2$ is only {\it multiplicatively} renormalised and so SM infrared (IR) contributions do not bring back the naturalness problem, once it has been eradicated from the UV. These considerations suggest a possible solution to the naturalness of the Higgs, which I will call here {\it UV Naturalness}. It is based on two assumptions: {\it (i)} a miracle occurs in quantum gravity, which sets $m_H^2(M_{\rm Pl})$ to be approximately zero ({\it i.e.} about 34 orders of magnitude smaller than the naive expectation); {\it (ii)} if there are new particles with mass between $M_{\rm Pl}$ and $m_h$, then they must be sufficiently decoupled from the Higgs field.  

In his {\it Summa contra gentiles}, St.~Thomas Aquinas classifies miracles in three categories. A miracle of the third degree is when God does something that nature can do, but without intervention of a natural agent ({\it e.g.} a storm that suddenly stops just before the ship sinks). A miracle of the second degree is when God does something that nature can do, but without respecting the natural temporal order ({\it e.g.} a man regains sight after being blinded or comes back to life after death). The highest degree of miracle is when God does something that nature can never do ({\it e.g.} parting the waters of the Red Sea or causing the sun to stand still at Gibeon). 

We can get inspiration from ancient wisdom and, in a modern {\it Summa contra naturalitatem}, classify the degree of quantum-gravity miracles required by the assumption {\it (i)} above. A miracle of the third degree occurs if graviton loops do not affect the Higgs mass and do not modify the evolution of the SM couplings in the far UV ({\it i.e.} in the transplanckian region). In this case gravity does not introduce a naturalness problem, but one may need to introduce new physics to avoid the non-asymptotic freedom of the hypercharge coupling or other possible Landau poles. A miracle of the second degree corresponds to a situation in which both gravity and the SM are well-behaved: the Higgs mass is not affected by any large corrections and all couplings reach UV fixed points. Finally, a first degree miracle would happen if quantum-gravity effects magically erase any large quantum correction to the Higgs mass generated at any scale, larger or smaller than $M_{\rm Pl}$. The latter possibility seems utterly implausible and I will disregard it, since it requires an exact correlation between contributions occurring at completely different energy scales. So, resorting to a quantum-gravity miracle (say of the second or third degree), we can conceive the possibility of a special boundary condition where $m_H^2(M_{\rm Pl})$ is about zero (in Planck units). In this context, it may well be that conformal symmetry plays a special role in ensuring the desired properties of the UV theory that determines the matching condition.

The requirement {\it (ii)} of UV Naturalness provides a formidable constraint on the theory. An extreme possibility, championed by Shaposhnikov~\cite{Shaposhnikov:2007nj}, is that the SM with the mere addition of very light right-handed neutrinos is everything there is in physics up to the Planck scale. Recently, it was pointed out~\cite{Farina:2013mla} that UV Naturalness allows for the existence of new physics at the weak scale ({\it e.g.} new dark-matter particles) or, in special circumstances, of even heavier particles. In particular, the see-saw mechanism can be consistent with UV Naturalness, as long as the right-handed neutrinos are lighter than $10^7$~GeV. Unfortunately, this bound is incompatible with the simplest scenarios of thermal leptogenesis. 

It is important to remark that constraints on new physics in UV Naturalness are different from the case of Unnaturalness. UV Naturalness mainly restricts the properties of heavy particles, while Unnaturalness restricts the properties of light particles. For instance, the discovery of a multi-Higgs structure at the weak scale is tolerated by UV Naturalness, while it would be unacceptable for Unnaturalness, where the lightness of a scalar particle is understood only in terms of anthropic considerations. It should also be pointed out that, from a theoretical point of view, UV Naturalness comes at a heavy price, since one has to give up many ideas (such as unification) based on high-scale physics. Moreover, unlike the case of the multiverse, UV Naturalness provides no clues regarding the cosmological constant.

\section*{IR Naturalness}

The common expectation is that new physics shuts off the Higgs mass sensitivity to quantum corrections below the TeV. I will refer to this case as {\it IR Naturalness}, since the naturalness problem is solved by dynamics occurring below the energy scale suggested by effective field theory arguments. For decades, IR Naturalness has stimulated revolutionary and inspiring ideas involving new physical concepts and symmetries. It would be a great triumph for science to discover that theories like supersymmetry, extra dimensions, compositeness, or technicolour describe the next layer of the micro-world, vindicating the concept of naturalness.  However, the LHC has undoubtedly put under considerable pressure the idea of IR Naturalness. To the pre-LHC constraints from electroweak data and rare processes, the LHC has added new bounds from direct searches, Higgs mass, Higgs couplings, and flavour-violating processes.

Take the case of supersymmetry. Under some simple assumptions (which may well represent a misleading oversimplification), the two parameters most relevant for naturalness -- the gluino and stop masses -- are excluded below 1.3-1.4 TeV and 600-700 GeV, respectively~\cite{oliver}. In the simplest setup, the Higgs mass gives an even more stringent constraint on stop masses, unless the stop mixing parameter takes a special (and extreme) value. Are these bounds problematic for IR Naturalness? A simple estimate~\cite{Barbieri:1987fn,Ellis:1986yg} can be obtained by considering the stop loop corrections to the physical Higgs mass $m_h^2$ ($=-2m_H^2$) 
\begin{equation}
\label{fine}
\frac{\delta m_h^2}{m_h^2} =\frac{3\, y_t^2\, {\tilde m}_t^2}{2\, \pi^2\, m_h^2}\ln \frac{\Lambda}{{\tilde m}_t}
\approx 100 \left( \frac{{\tilde m}_t}{600~{\rm GeV}}\right)^2 \left( \frac{\ln \Lambda / {\tilde m}_t}{30}\right).
\end{equation}
Roughly speaking, for a mediation scale around the unified scale ($\Lambda =10^{16}$~GeV), 
a stop of 600~GeV gives a contribution to $m_h^2$ which is 100 times larger than the physical value. This is an enormous improvement over the original fine-tuning problem $m_h^2/M_{\rm Pl}^2\approx 10^{-34}$, but it falls short of solving naturalness, leaving behind a tuning at the level of percent. After the results from LHC8, this level of tuning is generally present  even in the most optimistic case of split squark generations~\cite{Dimopoulos:1995mi,Cohen:1996vb}.

Equation~(\ref{fine}) quantifies the problem, but also suggests ways in which IR Naturalness could be saved. One possibility is to reduce the logarithmic running by taking smaller values of $\Lambda$. This is realised in models with low-mediation scale or in models where the Higgs mass is further protected ({\it e.g.} in supersoft supersymmetry with Dirac gauginos~\cite{Fox:2002bu}). Another possibility is to evade the direct limits on stop and gluino masses by considering compressed spectra~\cite{Martin:2007gf}, hadronic R-parity violation, or new decay chains~\cite{Fan:2011yu}. The indirect limits on stop masses from the Higgs mass measurement can be relaxed by extending the theory to include extra contributions to the Higgs quartic coupling, beyond the pure supersymmetric weak-gauge effect. This can be done with a new Higgs singlet (the so-called NMSSM), with new low-energy gauge groups~\cite{Batra:2003nj,Maloney:2004rc}, new vector-like fermions~\cite{Martin:2009bg,Graham:2009gy}, or new scalars with large quartic couplings and mixings with the Higgs~\cite{Galloway:2013dma}.

The situation about supersymmetry after LHC8 is far from settled, and some parameter regions with moderate fine-tuning still survive. However, in my opinion, the model-building complications required to achieve the reduction of the electroweak-scale tuning often come at the price of an increase of tuning in `theory space'. The situation is similar in the case of Higgs composite models, which are probably the most realistic competitors to supersymmetry as a solution to IR Naturalness~\cite{rob}. Even if a modest gain in the tuning of the electroweak condition can be achieved, the model-building difficulties related to a satisfactory explanation of flavour are a serious drawback. 

Experimental investigations of the Higgs couplings are direct probes of the naturalness of the Higgs boson (see {\it e.g.} refs.~\cite{Arvanitaki:2011ck,Craig:2013xia,Farina:2013ssa,Hedri:2013ina}). New physics that accounts for IR Naturalness must modify the Higgs self-energy and thus will inevitably affect Higgs vertices. It is then (almost) a theorem that the more natural the Higgs boson is, the more its properties must deviate from the Standard Model. 

Take the case of supersymmetry. Loops of stops modify the Higgs coupling to gluons and photons
\begin{equation}
\frac{\sigma (gg\to h)}{\sigma (gg\to h)_{\rm SM}}=\left( 1+\Delta_t\right)^2,~~~~
\frac{\Gamma (h \to \gamma \gamma)}{\Gamma (h \to \gamma \gamma)_{\rm SM}}=\left( 1-0.3\Delta_t\right)^2
\end{equation}
\begin{equation}
\Delta_t=\frac 14 \left( \frac{m_t^2}{{\tilde m}_{t_1}^2}+ \frac{m_t^2}{{\tilde m}_{t_2}^2}-\frac{m_t^4A_t^2}{{\tilde m}_{t_1}^2{\tilde m}_{t_2}^2}\right) \approx \left( \frac{600~{\rm GeV}}{{\tilde m}_t}\right)^2 4\% \ .
\end{equation}
For illustration, in the last equation I took degenerate stops (${\tilde m}_{t_1}={\tilde m}_{t_2}$) and no mixing ($A_t=0$). This shows that in a pre-LHC8 situation, when stop masses could be close to $m_t$, large deviations of the Higgs couplings from the SM prediction could be expected, in agreement with the `theorem' quoted above. For instance, for ${\tilde m}_{t}=200$~GeV, the product $\sigma (gg\to h)\Gamma (h \to \gamma \gamma)$ could differ from the SM by 50\%. The same product, for TeV stop masses, is modified only by 2\%. In other words, supersymmetry is becoming sufficiently `unnatural' to preserve the SM properties of the Higgs boson. This shows that `weakly-interacting' dynamics at the Fermi scale (such as supersymmetry), which affect the Higgs properties at the loop level, are best probed by direct searches rather than by precise determinations of Higgs couplings. The situation is different in the case of `strongly-interacting' sectors (such as composite Higgs) or tree-level modifications, where studies of Higgs couplings provide an efficient probe of new dynamics, competitive with direct bounds. 

\section*{Conclusions}

I made the point that naturalness is not an idle theoretical idea, but is a concept deeply rooted in our approach to physical phenomena based on effective-field theories. This does not mean that the principle of naturalness is necessarily valid in nature at all scales. So far it has been a successful guide for us to infer the energy scale at which a certain effective theory breaks down and a new physical description sets in.  
But perhaps we are reaching the stage at which our vision of the physical world as a stack of effective field theories is failing and a new picture is lurking behind. Whatever the truth is, there is no
doubt that testing the naturalness principle at the weak scale has far-reaching consequences for particle physics,  decisive for the future of our field.

Here I have reviewed various options. In the case of Unnaturalness, 
the most concrete known setup is the multiverse, which has the virtue of addressing both the Higgs and the cosmological constant problems, but the vice of possibly drawing physics away from experimental test.
However, even in the case of Unnaturalness, new physics around the weak scale is possible, although discoveries at the LHC are not guaranteed.
UV Naturalness deals with the situation in which there are no interactions other than gravity that can induce large quantum corrections to the Higgs mass. It can be viewed as an open possibility, although today it relies on unproven (and questionable) quantum-gravity miracles. Also in this case new physics is possible, but highly constrained.
Finally, the most welcome outcome for the future of experimental high-energy physics would be the discovery of new physics that accounts for the Higgs naturalness problem at the predicted scale. So far, the message from the LHC has not been encouraging, but the final verdict will have to wait for the high-energy run. The last word about the fate of naturalness is now mostly an experimental issue.

\bigskip
I want to thank A.~Arvanitaki, S.~Dimopoulos, S.~Dubovsky, G.~Dvali, J.~March-Russell, R.~Rattazzi, A.~Strumia, and G.~Villadoro for discussions on the ideas presented in this talk.

\end{document}